\documentclass[lettersize,journal]{IEEEtran}
\usepackage{graphicx}
\usepackage{multirow}
\usepackage{dblfloatfix} 
\usepackage[dvipsnames]{xcolor}
\usepackage{soul, url}
\usepackage{color}
\usepackage{boldline}
\usepackage{xcolor,colortbl}
\usepackage{listings}

\usepackage{tikz}

\usepackage{pifont}

\definecolor{Gray}{gray}{0.85}
\definecolor{LightCyan}{gray}{0.85}

\newcolumntype{g}{>{\columncolor{Gray}}c}

\usepackage[separate-uncertainty = true,multi-part-units = repeat]{siunitx}
\usepackage{array,amsfonts,amsmath}
\usepackage[font=small]{caption}
\usepackage{subcaption}

\usepackage{fancyhdr}
\usepackage{lipsum}

\fancypagestyle{header}{
    \fancyhf{}
    \fancyhead[C]{\textcolor{red}{This work has been accepted to the IEEE Computational Intelligence Magazine for possible publication. Copyright may be transferred without notice, after which this version may no longer be accessible. This is the initial version. The updated version can be downloaded from IEEE.}}
    
}

\begin{document}

\title{Can Large Language Models Aid in Annotating Speech Emotional Data? Uncovering New Frontiers}

\author{Siddique Latif, Muhammad Usama, Mohammad Ibrahim Malik,  and \\Bj\"{o}rn W.\ Schuller,~\IEEEmembership{Fellow,~IEEE}
\IEEEcompsocitemizethanks{

Corresponding E-mail: siddique.latif@qut.edu.au}}

\IEEEtitleabstractindextext{%
\begin{abstract}
Despite recent advancements in speech emotion recognition (SER) models, state-of-the-art deep learning (DL) approaches face the challenge of the limited availability of annotated data. Large language models (LLMs) have revolutionised our understanding of natural language, introducing emergent properties that broaden comprehension in language, speech, and vision. This paper examines the potential of LLMs to annotate abundant speech data, aiming to enhance the state-of-the-art in SER. We evaluate this capability across various settings using publicly available speech emotion classification datasets. Leveraging ChatGPT, we experimentally demonstrate the promising role of LLMs in speech emotion data annotation. Our evaluation encompasses single-shot and few-shots scenarios, revealing performance variability in SER. Notably, we achieve improved results through data augmentation, incorporating ChatGPT-annotated samples into existing datasets. Our work uncovers new frontiers in speech emotion classification, highlighting the increasing significance of LLMs in this field moving forward.

\end{abstract}

\begin{IEEEkeywords}
Speech emotion recognition, data annotation, data augmentation, large language models
\end{IEEEkeywords}}

\maketitle
\thispagestyle{header}
\IEEEdisplaynontitleabstractindextext
\IEEEpeerreviewmaketitle

\section{Introduction
\label{sec:introduction}}
\pagestyle{header}
The rapid growth in Natural Language Processing (NLP) has led to the development of advanced conversational tools, often called large language models (LLM) \cite{brown2020language}. These tools are capable of assisting users with various language-related tasks, such as question answering, semantic parsing, proverbs and grammar correction, arithmetic, code completion, general knowledge, reading comprehensions, summarisation, logical inferencing, common sense reasoning, pattern recognition, translation, dialogues, joke explanation, educational content, and language understanding \cite{brown2020language}. LLMs are trained on an enormous amount of general-purpose data and human-feedback-enabled reinforcement learning. A new field of study called ``Foundational Models" has emerged from these LLMs, highlighting the interest of the academic community and computing industry \cite{bommasani2021opportunities}. The foundational models have demonstrated the ability to perform tasks for which they were not explicitly trained. This ability, known as emergence, is considered an early spark of artificial general intelligence (AGI) \cite{wei2022emergent}. The emergence properties of the foundational models have sparked a wide range of testing of these models for various tasks, such as sentiment analysis, critical thinking skills, low-resource language learning and translation, sarcasm and joke understanding, classification, and other affective computing challenges. 

Speech emotion recognition (SER) is a fundamental problem in affective computing. The need for SER has evolved rapidly with the rapid integration of modern technologies in every aspect of our lives. SER systems are designed to understand the wide range of human emotions from the given input data (audio, video, text, or physiological signal) using traditional and modern machine learning (ML) techniques \cite{latif2023transformers,latif2022deep}. However, the availability of larger annotated data remains a challenging aspect for speech emotion recognition (SER) systems, which prompts the need for further investigation and exploration of new methods.

The use of crowd-sourced and expert intelligence for data annotation is a common practice. The annotated data serves as the ground truth for ML models to learn and generate predictions. This annotation policy is mostly opted in computational social science (sentiment analysis, bot detection, stance detection, emotion classification, etc.), human emotion understanding, and image classification \cite{cioffi2017computation,latif2020deep}. However, these strategies are prone to a variety of biases, ranging from human biases to situational biases \cite{latif2022ai,latif2019caveat}. These annotation techniques also necessitate a big pool of human annotators, clear and straightforward annotator instructions, and a verification rationale that is not always available or dependable \cite{rottger2021two}. Although there are a few unsupervised techniques for data annotations, these techniques necessitate a high sample size of the data; unfortunately, the generated annotations do not embed the context \cite{liao2019unsupervised}.  

Annotating speech emotion data is a doubly challenging process. The annotators listen to a speech recording and assign an annotation to a data sample using the pre-defined criteria. Human emotions are highly context-dependent, and annotating emotions based on a brief recording in a specific controlled situation might restrict the annotations' accuracy. Though the state-of-the-art on human-annotated emotion classification is strong, the generalisability of the learning for unseen data with slightly different circumstances might stymie the SER system's effectiveness. The recent availability of several LLMs (ChatGPT, Google Bard, etc.) has unearthed the possibility of replacing or assisting human annotators. LLMs are trained on enormous text corpora, allowing them to learn and grasp complicated language patterns. Their emergence property \cite{burns2022discovering} makes them well-suited for data annotations and various studies (e.\,g., \cite{zhu2023can,huang2023chatgpt}) explored LLMs for annotations of various natural language processing (NLP) tasks. However, none of the studies explores them to annotate speech emotion data based on the transcripts. 

In this paper, we present an evaluation of the effectiveness of large language models (LLMs) 
in annotating speech data for SER. We performed a series of experiments to show the effectiveness of ChatGPT for data annotation. However, we observed that annotations solely based on text lacked generalisation to speech emotion data due to the absence of audio context. To address this limitation, we propose a novel pipeline that incorporates audio features such as average energy, pitch, and gender information to provide essential audio context for accurate sample annotation. Furthermore, we introduce a method for encoding speech into a fixed-length discrete feature representation using a Vector Quantised Variational Autoencoder (VQ-VAE) \cite{ding2019group}, which serves as the audio context in the annotation prompt. To the best of our knowledge, this is the first endeavour to leverage LLMs for annotating speech emotion data, specifically for classification purposes, and evaluating their performance. We conduct a comparative analysis between LLM-based data annotations and human data annotations using publicly available datasets, including IEMOCAP and MSP-IMPROV. 

In the following section, we provide a brief literature review on the use of LLMs for data annotation. We highlight the gap between conventional annotations and annotations made with LLMs. Section III covers the methodology used in this study, Section IV presents the initial results and compares the performance of various LLMs for speech emotion data annotation, Section V provides a detailed discussion of the results and limitations, and Section VI concludes the paper with the potential to extend this work.

\section{Related Work}

This section provides an overview of the research on leveraging fundamental models such as LLMs for data annotation \cite{yang2023harnessing}. Data annotations are critical for developing ML models capable of uncovering complex patterns in large datasets and pushing the 
state-of-the-art 
in a particular domain. Human expert annotators, bulk annotations, semi-supervised annotations, and crowdsourced annotations are all widely used approaches in practice \cite{pustejovsky2012natural}. These strategies have their pros and cons. Human annotators, for example, can provide high-quality data annotations but are susceptible to challenges such as fairness, bias, subjectivity, high cost and time, label drifting, annotation fatigue and inconsistency, dealing with data ambiguity, and scalability. Bulk annotations are a faster and less expensive technique to create data annotations, but they might result in lower-quality annotations. Semi-supervised annotations combine the benefits of human-expert annotations with bulk annotations for data annotation, but they are complex to implement and have generalisability and robustness difficulties. Although crowdsourcing human intelligence to annotate large datasets is the quickest and most cost-effective option, it can create lower-quality annotations and is more challenging to manage the quality of the annotations. 

Recently, a few studies have investigated the efficacy of LLMs (i.\,e., ChatGPT) for data annotations. The goal of these experiments was to explore the potential of ChatGPT for data annotation and to find out whether ChatGPT can achieve full emergence in downstream tasks such as classification. Zhu et al.\ \cite{zhu2023can} tested the ability of ChatGPT to reproduce the human-generated annotations for five seminal computational social science datasets. The datasets include stance detection (two datasets), hate speech detection, sentiment analysis, and bot detection. Their results indicate that ChatGPT is capable of annotating the data, but its performance varies depending on the nature of the tasks, the version of ChatGPT, and the prompts. The average re-annotation performance is 60.9\% across all five datasets. For the sentiment analysis task, the accuracy of ChatGPT re-annotating the tweets is reported at 64.9\%, and for the hate speech task, the ChatGPT performance has gone down to 57.1\%. The authors also provided a prompt template that was used for re-annotating the data. 

Fact-checking is a well-known way to deal with the misinformation epidemic in computational social science. Hose et al.\ \cite{hoes2023using} evaluated the ability of LLMs, specifically ChatGPT, to assist fact-checkers in expediting misinformation detection. They used ChatGPT as a zero-shot classifier to re-annotate 12,784 human-annotated (``true claim", ``false claim") fact-checked statements. ChatGPT was able to correctly re-annotate 72.0\% of the statements. The study further suggests that ChatGPT performs well on recent fact-checked statements with ``true claim" annotations. 
Despite the reasonable performance of ChatGPT on fact-checking, it is hard to suggest that it will replace human fact-checkers anytime soon. Yang et al.\ \cite{yang2023large} explored the rating of news outlet credibility by formulating the problem as a binary re-annotation task for ChatGPT. ChatGPT achieved a reasonable performance in re-annotating 7,523 domains with a Spearman correlation coefficient of $\rho$ = 0.54. Tornberg \cite{tornberg2023chatgpt} also used ChatGPT-4 as a zero-shot classifier for re-annotating 500 political tweets. He found that ChatGPT-4 outperformed experts and crowd annotators in terms of accuracy, reliability, and bias. Gilardi et al.\ \cite{gilardi2023chatgpt} reported that ChatGPT used as a zero-shot classifier, outperformed the crowd-works-based text annotations for five text-annotation tasks around content moderation. We have also observed studies using LLMs (ChatGPT) for annotating/re-annotating data for various computational social science tasks such as election opinion mining tasks \cite{elmas2023opinion}, intent classification \cite{cegin2023chatgpt}, genre identification \cite{kuzman2023chatgpt}, stance detection \cite{mets2023automated}, and sentiment analysis \cite{wang2023chatgpt}. Several other prominent works that evaluate the application of LLMs in the annotation of computational social science datasets for various applications include \cite{ziems2023can, veselovsky2023generating, mu2023navigating, rytting2023towards}. 

Amin et al.\ \cite{amin38will} evaluated the capabilities of ChatGPT in three famous NLP classification tasks in affective computing: personality recognition, suicide tendency prediction, and sentiment analysis. Their results indicated that ChatGPT shows far better performance (in the presence of the noisy data) than Word2Vec models \cite{mikolov2013efficient}; ChatGPT further produces comparable performance with Bag-of-Words (BoW) and Word2Vec models (without noisy data) and was outperformed by a RoBERTa model \cite{liu2019roberta} trained for a specific affective computing task. ChatGPT scored an unweighted average recall of 85.5\% on the sentiment analysis, outperforming BoW and Word2Vec models by nearly 20.0\%. RoBERTa also scored an unweighted average recall of 85.0\%
on this task.
For the suicide tendency prediction task, ChatGPT's performance was the same as Word2Vec and BoW, with all three models achieving an unweighted average recall of nearly 91.0\%. RoBERTa outperformed ChatGPT on this task, achieving an unweighted average recall of 97.4\%. For the personality recognition task, RoBERTa performed best, scoring an unweighted average recall of 62.3\%. ChatGPT performed the worst on this task, getting an unweighted average recall of 54.0\%.
Interestingly, Word2Vec and BoW models also performed marginally well when compared to ChatGPT for this task. 

Wang et al.\ \cite{wang2021want} argued that GPT-3 can be a low-cost solution for the data annotations for downstream natural language understanding and generation tasks. This research evaluated the efficacy of augmenting human-annotated data with GPT-3 annotated data for improving the performance (language understanding and generation) in a constrained annotation budget. They tested their method on various language understanding and generation tasks, ranging from sentiment analysis, question answering, summarisation, text retrieval to textual entailment. They found that GPT-3 based annotations policy saved 50.0\% to 96.0\% cost in annotation tasks. However, they also noted that GPT-3 is not yet as reliable as human annotators in annotating high-stakes sensitive cases. More details on the evaluation of the comparison of ChatGPT with human experts on various NLP tasks are compared and evaluated in \cite{guo2023close}. Huang et al.\ \cite{huang2023chatgpt} explored the ability of ChatGPT to reproduce annotations and their corresponding natural language explanation. Their results indicate that lay people agreed with the results more when they were provided with the ChatGPT-generated natural language explanation of the annotations than just the considered post itself along with the annotation. ChatGPT agreed with the human-annotated data points 80.0\% of the time.

In contrast to the aforementioned studies, our research explores the untapped potential of LLMs in annotating emotions in speech data. We present a novel approach that incorporates audio context into LLMs to improve the precision of annotations. To our knowledge, no prior research has investigated the utilisation of LLMs for annotating speech emotion data.


\section{Methodology}

In our exploration of emotional data annotation, we conduct a series of experiments. Firstly, we annotate samples using only text, and then we incorporate audio features and gender information alongside textual data for improved annotation. To incorporate audio context, we utilise the average energy and pitch of each utterance and pass it to ChatGPT. Additionally, we propose the use of VQ-VAE to generate a 64-dimensional discrete representation of audio, which is also provided to ChatGPT as the audio context. For speech-emotion classification, we train a bi-directional Long-Short Term Memory (BLSTM)-based classifier. The following section provides further details on our proposed method.



\subsection{VQ-VAE for Speech Code Generation}
\label{VQVAE}
\begin{figure}[!ht]
\centering
\includegraphics[width=0.5\textwidth]{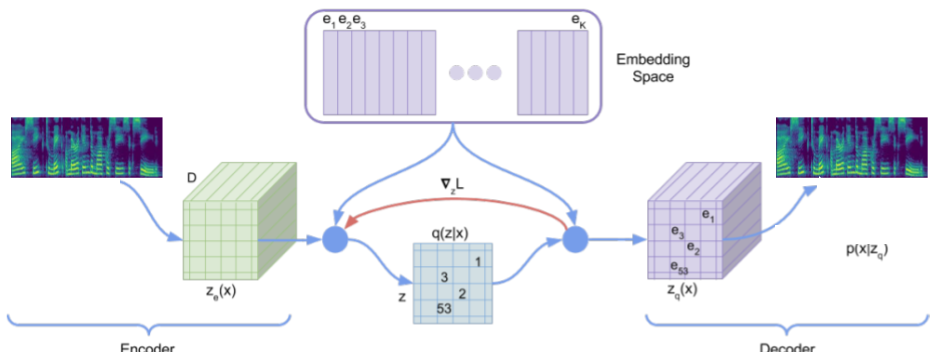}
\caption{Model Diagram of the VQ-VAE}
\label{fig:vqvae}
\end{figure}

We propose to use a Vector-Quantised Variational Autoencoder (VQ-VAE) \cite{oord2018neural} to learn a discrete representation from the speech data. Unlike traditional VAEs where the discrete space is continuous, VQ-VAEs express the latent space as a set of discrete latent codes and the prior is learnt rather than being fixed. As illustrated in Figure \ref{fig:vqvae}. the model is comprised of three main parts: the encoder, the vector quantiser, and the decoder. 

The encoder takes in the input in the form of Mel-spectrograms and passes it through a series of convolutional layers having a shape of $(n,h,w,d)$ where $n$ is the batch size, $h$ is the height, $w$ is the width and d represents the total number of filters after convolutions. Let us denote the output from the encoder as $z_e$. The vector quantiser component contains an embedding space with $k$ total vectors each with dimension $d$. The main goal of this component is to output a series of embedding vectors that we call $z_q$. To accomplish this, we first reshape $z_e$ in the form of $(n*h*w, d)$ and calculate the distance for each of these vectors with the vectors in the embedding dictionary. For each of the $n*h*w$ vectors, we find the closest of the $k$ vectors from the embedding space and index the closest vector from the embedding space for each $n*h*w$ vector. The discrete indices of each of the vectors in the embedding space are called codes, and we get a unique series of codes for each input to the model. The selected vectors are then reshaped back to match the shape of $z_e$. Finally, the reshaped vector embeddings are passed through a series of transpose convolutions to reconstruct the original input Mel-spectrogram. One problem with this approach is that the process of selecting vectors is not differentiable. To tackle this problem, the authors
simply copy the gradients from $z_q$ to $z_e$. 

The total loss is composed of three loss elements: the reconstruction loss, the code book loss, and the commitment loss. The reconstruction loss is responsible for optimising the encoder and decoder and is represented by:

\begin{equation}
 \text{Reconstruction\, Loss} = -log( p(x|z_q) ).    
\end{equation}
We use a code book loss which forces the vector embeddings to move closer to the encoder output $z_e$.\\
\begin{equation}
   \text{Code\, Book\, Loss} = ||sg[z_e(x)] - e||^2 , 
\end{equation}
where $sg$ is the stop gradient operator, this essentially freezes all gradient flows. $e$ are the vector embeddings and $x$ is the input to the encoder. And finally, for making sure that the encoder commits to an embedding we add a commitment loss.

\begin{equation}
   \text{Commitment\, Loss} = \beta ||z_e(x) - sg[e]||^2, 
\end{equation}
here $\beta$ is a hyperparameter that controls the weight we want to assign to the commitment loss.

Overall, we train the VQ-VAE model to represent the audio representation in the form of a discrete list of integers or ``codes". These audio representations can be used in addition to the transcriptions and fed to ChatGPT for annotation. In the following section, we will delve into the details of the annotation procedure.

\subsection{Emotion Label Annotation using LLMs}
\label{emotionlabels}
We evaluated the data annotation ability of ChatGPT with different experiments. We start our experiments by annotating the training data of IEMOCAP by passing the textual transcripts to ChatGPT and annotating the data both in zero-shot and few-shot settings. For a few shots, we randomly selected 10 samples from the training data and passed them to ChatGPT as context. We trained the classifier using the training samples annotated with ChatGPT and unweighted average recall (UAR) is computed. We repeat this procedure of annotation by passing the audio features along with the textual information. First of all, we 
use average pitch and energy for a given utterance and re-annotated the data both in a zero-shot and a few-shots setting, and classification UAR is measured using a BLSTM based classifier. As the female voice usually has a high pitch and energy, therefore, we also annotated the data by providing the gender information. Finally, we propose to use an audio representation by VQ-VAE (Section \ref{VQVAE}) and pass it to ChatGPT as audio context. We then used the OpenAI API with the ``ChatGPT pro" version to annotate the data. In our approach, we meticulously designed and curated multiple prompts for annotating the data, leveraging ChatGPT for the annotation process. We trained the classifier on the annotated dataset and computed the UAR, considering it as a benchmark for evaluating the classification performance. To improve upon this benchmark, we conducted additional experiments, exploring various prompts to enhance the classification results beyond the established performance level.

\subsection{Speech Emotion Classifier}
In this work, we implement convolutional neural network (CNN)-BLSTM-based classifiers due to their popularity in SER research \cite{latif2021survey}. It has been found that the performance of BLSTM can be improved by feeding it with a good emotional representation \cite{trigeorgis2016adieu}. Therefore, we use CNN as emotional feature extractor from the given input data \cite{latif2019direct}. A CNN layer acts like data-driven filter banks and can model emotionally salient features. We pass these emotional features to the BLSTM layer to learn contextual information. Emotions in speech are in the temporal dimension, therefore, the BLSTM layer helps model these temporal relationships \cite{qayyum2018quran}. We pass the outputs of BLSTM to an attention layer to aggregate the emotional salient attributes distributed over the given utterance. For a given output sequence $h_{i}$, utterance level salient attributes are aggregated  as follows:
\begin{equation}
    R_{\text{attentive}}=\sum_{i}\alpha_{i}h_{i},
\end{equation}
where $\alpha_{i}$ represents the attention weights that can be computed as follows:
\begin{equation}
    \alpha_{i}=\frac{\text{exp}W^T h_{i}}{\sum_{j}\text{exp}W^T h_{j}},
\end{equation}
where $W$ is a trainable parameter. The attentive representation $R_{\text{attentive}}$ computed by the attention layer is passed to the fully connected layer for emotion classification. Overall, our classifier is jointly empowered by the CNN layers to capture an abstract representation, the BLSTM layer for context capturing, the attention layer for emotional salient attributes aggregation, and the fully connected layer emotion classification.

\section{Experimental Setup}

\subsection{Datasets}
\label{sec:data}
To evaluate the effectiveness of annotations by ChatGPT, we use three datasets: IEMOCAP, MSP-IMPROV, and MELD which are commonly used for speech emotion classification research \cite{Lotfian+2016,kim2016emotion}. Both, the IEMOCAP and the MSP-IMPROV datasets are collected by simulating naturalistic dyadic interactions among professional actors and have similar labelling schemes. MELD contains utterances from the Friends TV series.

\subsubsection{IEMOCAP}
  The Interactive Emotional Dyadic Motion Capture (IEMOCAP) database is a multimodal database that contains 12 hours of recorded data \cite{busso2008iemocap}. The recordings were captured during dyadic interactions between five male and five female speakers. The Dyadic interactions enabled the speakers to converse in unrehearsed emotions as opposed to reading from a text. The interactions are almost five minutes long and are segregated into smaller utterances based on sentences, where each utterance is then assigned a label according to the emotion. Overall, the dataset contains nine different emotions. To be consistent with previous studies, we use four emotions including sad (1084), happy (1636), angry (1103), and neutral (1708).


\subsubsection{MSP-IMPROV}
This corpus is a multimodal emotional database recorded from $12$ actors performing dyadic interactions \cite{busso2017msp}, similar to IEMOCAP \cite{busso2008iemocap}. The utterances in MSP-IMPROV are grouped into six sessions, and each session has recordings of one male and one female actor. The scenarios were carefully designed to promote naturalness while maintaining control over lexical and emotional contents. The emotional labels were collected through perceptual evaluations using crowdsourcing \cite{burmania2016increasing}. The utterances in this corpus are annotated in four categorical emotions: angry, happy, neutral, and sad. To be consistent with previous studies \cite{latif2019direct,gideon2017progressive},  we use all utterances with four emotions: anger (792), sad (885), neutral (3477), and happy (2644). 

\subsubsection{MELD}
Multimodal EmotionLines Dataset \cite{poria-etal-2019-meld} or MELD contains over 1400 dialogues and 13000 utterances and multiple speakers from the popular TV series Friends. The utterances have been labelled from a total of seven emotions: Anger, Disgust, Sadness, Joy, Neutral, Surprise and Fear. Furthermore, MELD also contains sentiment annotations for each utterance. To stay consistent with the other datasets we choose four emotions including sadness (1002 samples), neutral (6436 samples), joy and anger (1607 samples). With this configuration, we get a total of 11353 utterances from the dataset.  


\subsection{Speech Features}
For utterances across all datasets, we use a consistent sampling rate of 16\,kHz. For extracting the audio features we then convert the audio into Mel spectrograms. The Mel-spectrograms are computed with a short-time Fourier transform of size 1024, a hop size of 256, and a window size of 1024. We specify a total of 80 Mel-bands for the output and cutoff frequency of 8\,kHz. We set a cutoff length of 256 for each Mel spectrogram to have a final shape of 80x256, where smaller samples are zero-padded. Finally, the Mel spectrograms are normalised in the range of $[-1,1]$.

\subsection{Hyperparameters}

The VQ-VAE was trained using the following parameters: We chose a batch size of 256 and trained for a total of 1000 epochs with a learning rate of $1e^{-4}$. The convolution layers each had a stride and kernel size of 2 and 3, respectively. A total of 8192 token embeddings were selected, where each had a dimensionality of 512. With our particular configuration, we got a total of 64 codes for each given utterance. We pass these codes to ChatGPT along with textual data for annotation. Based on these annotations, we trained over the classifier.

Our classifier consists of convolutional layers and a Bidirectional LSTM (BLSTM)-based classification network. To generate high-level abstract feature representations, we employ two CNN layers. In line with previous studies \cite{dai2019learning,gideon2019improving}, we utilise a larger kernel size for the first convolutional layer and a smaller kernel size for the second layer. The CNN layers learn feature representations, which are then passed to the BLSTM layer with 128 LSTM units for contextual representation learning. Following the BLSTM layer, an attention layer is applied to aggregate the emotional content spread across different parts of the given utterance. The resulting attentive features are then fed into a dense layer with 128 hidden units to extract emotionally discriminative features for a softmax layer. The softmax layer employs the cross-entropy loss function to calculate posterior class probabilities, enabling the network to learn distinct features and perform accurate emotion classification.

In our experiments, we utilise the Adam optimiser with its default parameters. The training of our models starts with a learning rate of $0.0001$, and at the end of each epoch, we assess the validation accuracy. If the validation accuracy fails to improve for five consecutive epochs, we decrease the learning rate by half and revert the model to the best-performing previous epoch. This process continues until the learning rate drops below $0.00001$. As for the choice of non-linear activation function, we use the rectified linear unit (ReLU) due to its superior performance compared to leaky ReLU and hyperbolic tangent during the validation phase.

\section{Experiments and Results}
\label{exper}

All experiments are conducted in a speaker-independent manner to ensure the generalisability of our findings. Specifically, we adopt an easily reproducible and widely used leave-one-speaker-out cross-validation scheme, as commonly employed in related literature \cite{latif2018variational,latif2020multi,bao2019cyclegan}. For cross-corpus SER, we follow \cite{bao2019cyclegan,latif2022multitask} and use IEMOCAP for training and MSP-IMPROV is used for validation and testing. For the experiments, we repeat each experiment ten times and calculate the mean and standard deviation of the results.  The performance is presented in terms of the unweighted average recall rate 
(UAR), a widely accepted metric in the field that more accurately reflects the classification accuracy across multiple emotion categories 
when the data is in imbalance across these.

\subsection{Within Corpus Experiments} 
\label{within}
\begin{figure}[!ht]
\centering
\includegraphics[width=0.46\textwidth]{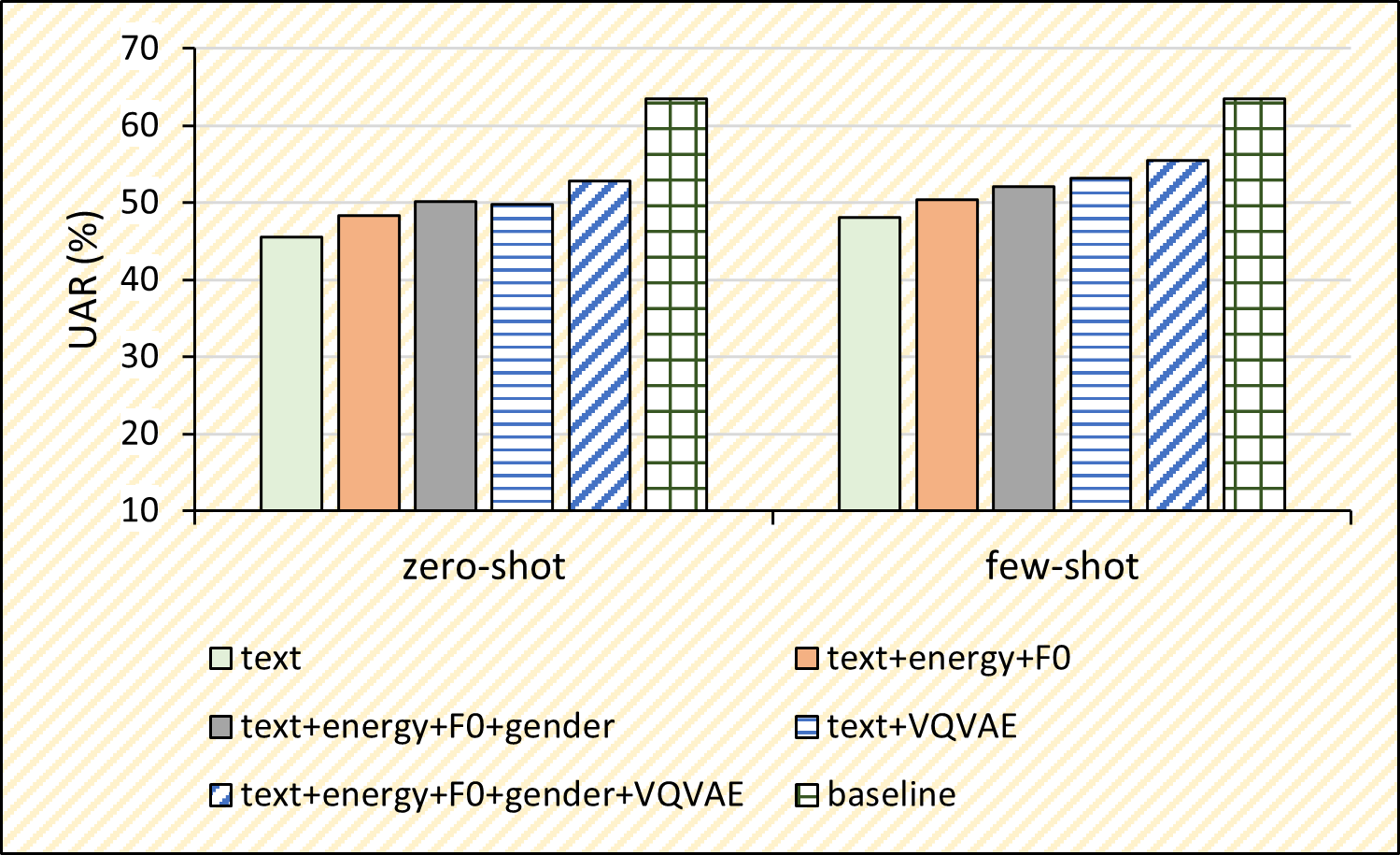}
\caption{Comparing the classification performance (UAR \%) using training data annotated by ChatGPT and original IEMOCAP labels.}
\label{fig:within_corpus}
\end{figure}

For the within-corpus experiments, we select the IEMOCAP data and compare the results with the baseline UAR achieved using actual true labels. 
We trained the classifier for different settings: (1) true label settings, (2) zero-shot ChatGPT labels, and (3) few-shots ChatGPT labels. In the first experiment, we trained the CNN-BSTM-based classifier on true labels using the well-known above mentioned leave-one-speaker-out scheme \cite{malik2023preliminary,latif2020deep}. In the second and third experiments, the classifier is trained in the same leave-one-speaker-out scheme, however, we annotated samples using ChatGPT with our proposed approach. We repeat the second and third experiments using text only and text plus audio context. Results are presented in Figure \ref{fig:within_corpus}. Overall, results on data annotated using few shots achieve improved results compared to the zero-shot scenario. 
It is important to note that the emotion classification performance using training data annotated with only text is poor compared to the baseline. Here, baseline results represent when the classifier is trained using the original annotations of IEMOCAP. This observation underscores the insufficiency of textual information alone to provide the necessary context for accurate annotation by ChatGPT. Consequently, additional context becomes essential to enable ChatGPT in effectively annotating the data. 
As previously found, for example, happy and angry voice samples often have high energy and pitch compared to a sad and neutral voice \cite{yildirim2004acoustic}. Building upon this insight, we incorporated the average energy and pitch values of a given utterance as additional contextual information for ChatGPT during the re-annotation process, both in zero-shot and few-shot settings. 
However, the performance improvement was not considerable,
primarily due to the confounding factor of gender, as female voices typically exhibit higher pitch and energy compared to male voices \cite{fraccaro2011experimental}. To address this limitation, we extended the experiment by providing gender labels to ChatGPT, resulting in improved classification accuracy as illustrated in \ref{fig:within_corpus}. In addition to average energy, pitch, and gender information, we further proposed the utilisation of audio patterns to provide enhanced audio context for annotation. To achieve this, we employed a VQ-VAE model to encode the given utterance into discrete representations. These representations, along with the textual and other feature inputs, were employed in various experiments for annotation (refer to Figure \ref{fig:within_corpus}). Notably, in the zero-shot scenario, no substantial improvements were observed. However, significant
advancements were achieved by incorporating the discrete codes generated by VQ-VAE, in conjunction with average energy, pitch, and gender information.

\subsection{Cross-Corpus Evaluations}
\label{cross-corpus}
In this experiment, we perform a cross-corpus analysis to assess the generalisability of annotations performed using our proposed approach. Here, we trained models on IEMOCAP, and testing is performed on the MSP-IMPROV data. IEMOCAP is more blanched data, therefore, we select it for training by following previous studies ~\cite{latif2020multi,neumann2019improving,sahu2018enhancing}. We randomly select 30.0\,\% of the MSP-IMPROV data for parameter tuning and 70.0\,\% of data as testing data. 
We report results using the few-shots annotation by ChatGPT as it consistently demonstrated superior performance compared to the zero-shot setting.
\begin{table}[!ht]
\centering
\caption{Cross-corpus evaluation results for speech emotion recognition. }
\begin{tabular}{|l|l|}
\hline
Model                                         & UAR (\%)\\ \hline
\begin{tabular}[c]{@{}l@{}}Attentive CNN \cite{neumann2019improving}\end{tabular} &45.7\\ \hline
\begin{tabular}[c]{@{}l@{}}CNN-BLSTM$_{\scriptsize{(\text{baseline})}}$ \end{tabular} & 45.4$\pm$0.83  \\ \hline
\begin{tabular}[c]{@{}l@{}}text+energy+f0+gender\end{tabular} & \textbf{41.5$\pm$1.2 }\\ \hline
\begin{tabular}[c]{@{}l@{}}text+energy+f0+gender+VQ-VAE \end{tabular} & \textbf{42.7$\pm$0.9 }\\ \hline
\end{tabular}
\label{cross}
\end{table}

We compare our results with different studies in Table \ref{cross}. In \cite{latif2022multitask}, the authors use the CNN-LSTM model for cross-corpus evaluation. They show that CNN-LSTM can learn emotional contexts and help achieve improved results for cross-corpus SER. In \cite{neumann2019improving}, the authors utilise the representations learnt from unlabelled data and feed it to an attention-based CNN classifier. They show that the classifier's performance can be improved by augmenting the classifier with information from unlabelled data. We compare our results using the CNN-BLSTM-based classifier by using the IEMOCAP annotated by the ChatGPT model. This experiment demonstrates the generalisability of annotations performed by ChatGPT in cross-corpus settings. However, it is worth noting that our results did not surpass those of previous studies. In the subsequent experiment, we aim to showcase the potential for enhancing the performance of SER using data annotations generated by ChatGPT, both within-corpus and cross-corpus settings.

\subsection{Augmentating the Training Data}
\label{augment}
\begin{figure}[!ht]
\centering
\includegraphics[width=0.45\textwidth]{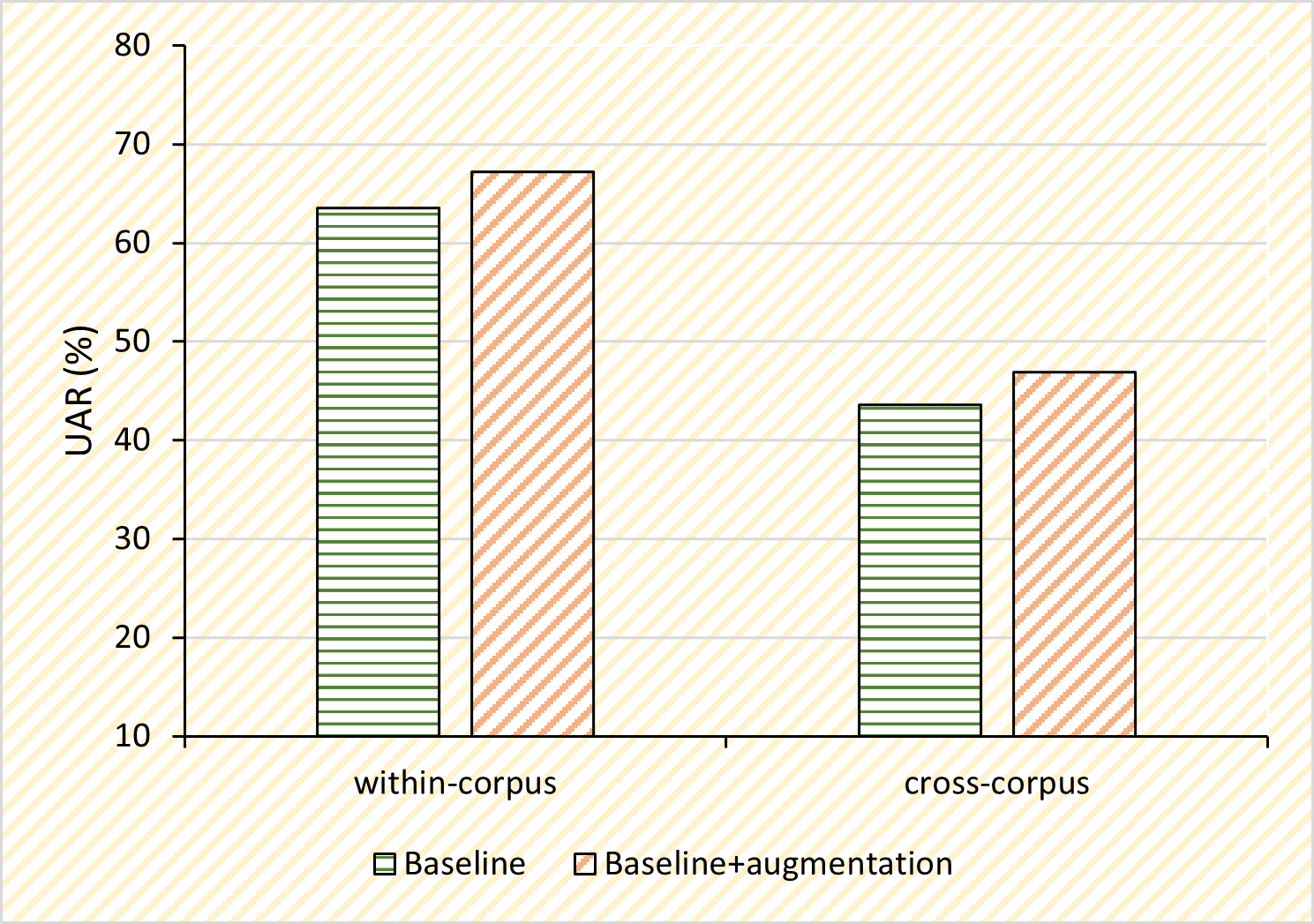}
\caption{Comparing the classier performance (UAR \%) with data augmentation.}
\label{fig:augmenting}
\end{figure}

In the previous two experiments, we showed, how we can annotate new speech-emotional data using a large language model like ChatGPT. However, the performance does not surpass the UAR achieved using actual labels. In this experiment, we aim to address this limitation by showcasing the potential of improving SER performance through data augmentation using our proposed approach.  For this, we can utilise abundantly available audio data by annotating with our proposed approach. For instance, data from YouTube can be annotated and used to augment the SER system. To validate this concept, we select the MELD dataset, which consists of dialogue samples from the Friends TV series. We employ the few-shot approach, using samples from the IEMOCAP dataset for few-shots, and annotate the MELD data with four emotions: happy, anger, neutral, and sad. We used samples from IEMOCAP data for the few-shots and annotated MELD data in four emotions including happy, anger, neutral, and sad. Results are presented in Figure \ref{fig:augmenting}, where we compare the results with the CNN-BLSTM classifier using the actual IECMOAP labels and when data is augmented using the samples with ChatGPT labels.
This analysis provides insights into the effectiveness of data augmentation for enhancing the performance of the SER system.

\begin{table}[!ht]
\centering
\caption{Comparison of results with previous studies.}
\begin{tabular}{|ll|}
\hline
\multicolumn{1}{|l|}{Model}     & UAR (\%) \\ \hline
\multicolumn{2}{|c|}{within corpus}     \\ \hline
\multicolumn{1}{|l|}{DialogueRNN \cite{majumder2019dialoguernn} (2019)}  & 63.40         \\ \hline
\multicolumn{1}{|l|}{CNN-attention \cite{peng2021efficient} (2021)} &  65.4        \\ \hline
\multicolumn{1}{|l|}{CNN-BLSTM (+ augmentation) (2022) \cite{latif2022multitask}}  &   65.1$\pm$1.8       \\ \hline

\multicolumn{1}{|l|}{Our work (+ augmentations) (2023)}   & \textbf{68.0$\pm$ 1.4 }      \\ \hline
\multicolumn{2}{|c|}{cross-corpus}         \\ \hline

\multicolumn{1}{|l|}{Cyclegan-DNN \cite{bao2019cyclegan} (+ augmentations) (2019)}  &   46.52$\pm$0.43
\\ \hline
\multicolumn{1}{|l|}{CNN-BLSTM (+ augmentations) \cite{latif2022multitask} (2022) }  &   46.2 $\pm$ 1.3\\ \hline
\multicolumn{1}{|l|}{Our work (+ augmentations) (2023)  }  &   \textbf{48.1$\pm$ 0.9}
\\ \hline

\end{tabular}
\label{table:augmenting}
\end{table}

Furthermore, we provide a comprehensive comparison of our results with previous studies in both within-corpus and cross-corpus settings, as presented in Table \ref{table:augmenting}. 
In \cite{majumder2019dialoguernn,peng2021efficient}, the authors utilise DialogueRNN for speech emotion recognition using IEMOCAP data. Peng et al.\ \cite{peng2021efficient} use an attention-based CNN network for emotion classification. We achieve better results compared to these studies by augmenting the classifier with additional data annotated by ChatGPT. One possible reason can be that these studies did not train the models with augmentation. However, we also compared the results with \cite{latif2022multitask}, where the authors use different data augmentation techniques to augment the classifier and achieve improved results. In contrast, we use ChatGPT to annotate the publicly available data and use it for augmentation of the training set. We are achieving considerably improved results compared to \cite{latif2022multitask}. One possible reason is that we are adding new data in the classifiers' training set, however, authors in \cite{latif2022multitask} employed perturbed versions of the same data, which can potentially lead to overfitting of the system. Similarly, we achieve considerably improved results for cross-corpus settings compared to the precious studies \cite{latif2022multitask,bao2019cyclegan}, where the authors augmented their classification models with either synthetic data or perturbed samples using audio-based data augmentation techniques like speed perturbation, SpecAugmet, and mixup.

Overall, our results showcase the effectiveness of our approach in achieving superior performance compared to previous studies, both in within-corpus and cross-corpus settings. The utilisation of ChatGPT for data annotation and augmentation proves to be a promising strategy for enhancing SER systems.

 \subsection{Limitations}
In this section, we highlight the potential limitations of our work and in general the limitations of LLMs for data annotation. During our experiments, we observed the following limitations:
\begin{itemize}
 \item We obtained promising results by augmenting the training data with samples annotated using ChatGPT. However, this approach proved ineffective when applied to corpora such as LibriSpeech \cite{panayotov2015librispeech}, where the recordings lack emotional variations. Although we attempted to utilise 
 LibriSpeech data (results are not shown here), the results were not as promising as those achieved with MELD. 
    \item ChatGPT is known to be sensitive to prompt variability, which can lead to ambiguous and erroneous results if even slight changes are made to the prompt content. In order to address this issue, we suggest conducting experiments using different prompts to generate annotations (as presented in Section \ref{emotionlabels}). The inclusion of more context in the prompts has been shown to improve the quality of results. However, for SER annotation prompts, this can be particularly challenging due to the significant variability of human emotions within short time frames. This limitation stems from LLMs' reliance on training data.
   
    \item ChatGPT has not been trained particularly to annotate speech emotion data. While the emergent nature of ChatGPT has aided with annotation, relying exclusively on ChatGPT annotation is insufficient. Through our research, we have found that incorporating ChatGPT-based annotations alongside the training data leads to enhanced classification performance. Notably, when utilising multi-shot ChatGPT annotations instead of zero-shot annotations, we observe a substantial performance improvement.
    \item ChatGPT offers a significant cost reduction in data annotation. For instance, in our experiments, we were able to annotate IEMOCAP data examples using ChatGPT for approximately 30 USD, which is significantly lower than human annotations cost. However, it is paramount to note that the accuracy of ChatGPT-based annotations is not as good as human annotations because ChatGPT is not specifically trained for annotating speech emotion data. As a result, it is a trade-off situation. Therefore, it becomes a trade-off between cost and accuracy. Striking the right balance is crucial when utilising ChatGPT for data annotation to avoid potential inaccuracies in classification performance.
\end{itemize}

Despite the mentioned limitations, we have found ChatGPT to be an invaluable tool for speech-emotion data annotation. We believe that its capabilities will continue to evolve. Currently, generating annotations using ChatGPT and incorporating them to augment human-annotated data has demonstrated improved performance in speech emotion classification. This highlights the potential of ChatGPT as a valuable asset in advancing research in this field.

\section{Conclusions and Outlook}

In this paper, we conducted a comprehensive evaluation of ChatGPT's effectiveness in annotating speech emotion data. To the best of our knowledge, this study is the first of its kind to explore the capabilities of ChatGPT in the domain of speech emotion recognition. The results of our investigation have been encouraging, and we have discovered promising outcomes. Below are the key findings of our study:
\begin{itemize}
    \item Based on our findings, we observed that text-based emotional annotations do not generalise effectively to speech data. To address this limitation, we introduced a novel approach that harnesses the audio context in annotating speech data, leveraging the capabilities of a large language model. By incorporating the audio context, we successfully enhanced the performance of SER, yielding improved results compared to the text-based approach.
    \item We observed that the quality of annotations by ChatGPT considerably 
    improved when using a few-shot approach compared to a zero-shot one. By incorporating a small number of annotated samples, we were able to achieve improved results in our evaluation.
    
    \item We introduced an effective technique to utilise large language models (LLMs) to augment the speech emotion recognition (SER) system with the annotated data by ChatGPT. The augmented system yielded improved results compared to the current state-of-the-art SER systems that utilise conventional augmentation techniques.  
     
\end{itemize}

In our future work, we aim to expand our experimentation by applying our approach to new datasets and diverse contexts. This will allow us to further validate the effectiveness and generalisability of our proposed technique. Additionally, we plan to explore and compare the annotation abilities of different LLMs for speech emotion data, enabling us to gain insights into their respective strengths and weaknesses. We also intend to use LLMs in the training pipeline of the SER system.

\end{document}